\documentclass[12pt]{article}
\usepackage{graphicx}
\usepackage{amssymb,amsmath,amsfonts,palatino,amsthm}
\usepackage{amssymb}
\usepackage{epstopdf}
\newcommand{\beq}{\begin{equation}}
\newcommand{\eeq}{\end{equation}}

\newcommand{\f}{\begin{equation}}
\newcommand{\ff}{\end{equation}}

\begin{document}

%%%%%%%%%%%%%%%%%%%%%%%%%%%%%%%%%%%%%%%%%%%%%%%%
\title{Unimodular loop quantum gravity and the problems of time}
\author{Lee Smolin\thanks{lsmolin@perimeterinstitute.ca}
\\
\\
Perimeter Institute for Theoretical Physics,\\
31 Caroline Street North, Waterloo, Ontario N2J 2Y5, Canada}
\date{\today}
\maketitle

\begin{abstract}
We develop the quantization of unimodular gravity in the Plebanski and Ashtekar formulations and
show that the quantum effective action defined by a formal path integral is unimodular.  
This means that the effective quantum geometry does not couple to terms in the expectation value of energy proportional to the metric tensor.  The path integral takes the same form as is used to define spin foam models, with the additional constraint that
the determinant of the four metric is constrained to be a constant by a gauge fixing term.  This extends the results of \cite{me-uni} to the Hilbert space and path integral of loop quantum gravity. 

We review the proposal of Unruh, Wald and Sorkin- that the hamiltonian quantization yields quantum evolution in a physical time variable equal to elapsed four volume-and discuss how this may be carried out 
in loop quantum gravity.

\end{abstract}
\newpage

\tableofcontents

\section{Introduction}

The unimodular formulation of general relativity was first proposed by Einstein in 1919 as an approach to the unification of gravity and matter\cite{einsteinuni} .  It was studied by a number of authors in the 1980s and early 90s because of indications that it resolves two key problems 
in quantum gravity\cite{zeeuni}-\cite{bombelliuni}.  These are the cosmological constant problem and the problem of defining a physically meaningful time with which to measure evolution of quantum states in quantum cosmology, in the absence of a spatial boundary.  This is the second of two papers which report results which support and clarify the sense in which unimodular
quantum gravity solves these two problems.

In the first of these papers\cite{me-uni}, I constructed the constrained phase space quantization of a formulation
of unimodular gravity due to Henneaux and Teitelboim\cite{HTuni}. I showed that the quantum effective action is a functional of the unimodular spacetime metric $g_{\mu \nu}$ with determinant fixed to
\f
\sqrt{det(\bar{g}_{\mu \nu})}=\epsilon_0
\label{uni1}
\ff
where $\epsilon_0$ is a fixed nondynamical volume element.  This means that the quantum effective
equations of motion, which arise from varying the metric with (\ref{uni1}) fixed have a symmetry 
\f
T_{ab} \rightarrow T^\prime_{ab} = T_{ab}+ \bar{g}_{ab} C
\label{modify}
\ff
where $C$ is a spacetime constant. This decouples the dynamics of the metric $\bar{g}_{}$ from any contribution to the energy-momentum tensor, whether classical or quantum, of the form of a constant times the spacetime metric.
This means that the puzzle of why huge contributions to $T_{\mu\nu}$ of this form coming from the zero point energy of the fields, expected to be of the form of $M_{Pl}^4$ are not sources of spacetime curvature.  This is sometimes called the old cosmological constant problem.

In this paper these results are rederived using the Ashtekar variables and the Plebanski form of the action. 
Specifically, in sections 2 to 4 we develop the unimodular formulation of the Ashtekar variables first studied  in \cite{bombelliuni}.  
In section 5 we formally construct the path integral and show that the quantum effective equations of motion have the unimodular property
(\ref{modify}).

This sets up the possibility of explicitly realizing unimodular quantum gravity in the 
context of a spin foam model, which is a well defined and ultraviolet finite path integral quantization of general relativity.  

In section 6 we then seek to build on these results to address the problem of time in quantum cosmology and the related issues of defining physical observables and the physical inner product.  Here we are following a suggestion made by Unruh\cite{unruhuni}, Wald\cite{unruhwald}, Sorkin\cite{sorkinuni} and others, which can be summarized as follows.  We define,
following Henneaux and Teitleboim a three form $a_{\alpha \beta \gamma}$ which satisfies a field equation
\f
\tilde{b}= (da)^* = \sqrt{-g}
\ff
Here $g_{\mu \nu}$ is the usual spacetime metric without the unimodular condition imposed. Let us consider a spatially compact spacetime $\cal M$.  On it we pick two
nonintersecting spacelike slices of a spacetime history , which we may call $\Sigma_2$ and $\Sigma_1$. These bound a region of spacetime $\cal R$.    It follows
that
\f
\int_{\Sigma_2} a -\int_{\Sigma_2} a = \int_{\cal R} \sqrt{-g} = Vol^4({\cal R})
\ff
We may then call 
\f
\tau (\Sigma) = \int_{\Sigma} a
\ff
the elapsed spacetime four volume to the past of the surface $\Sigma$.  

The idea of \cite{unruhuni,unruhwald,sorkinuni} is that the Hamiltonian constraints of general relativity are in this formulation replaced by
evolution equations in this physically well defined time coordinate.  To see explicitly what this means
define the usual $3+1$ slicing of spacetime, and set up a hamiltonian formulation of unimodular general relativity.  In each surface $\Sigma$ we can define a local time coordinate $\tilde{T}= a^*$.  This has  conjugate momenta,
$\pi$, so that 
\f
\{\tilde{T} (x) , \pi (y) \} = \delta^3 (y,x)
\label{A}
\ff
The hamiltonian constraints are then replaced by two sets of constraints.
\f
{\cal W}= \pi \sqrt{q} - \tilde{h} =0
\label{eq1}
\ff
where $\tilde{h}$ is the usual hamiltonian constraint, with vanishing cosmological constant, $\Lambda$, with density 
weight one.  The second set of constraints are
\f
{\cal G}_a = \partial_a \pi =0
\label{eq2}
\ff
This tells us that $\pi$ is a constant, and there are indeed equations of motion that confirm that it is a spacetime constant.  Its value on any solution is the cosmological constant.  From (\ref{A}) we see that this variable cosmological constant is the momentum conjugate to elapsed four volume.  

In section 6 we discuss the implementation of (\ref{eq1},\ref{eq2}) in an extension of the spin network representation.  We show that the Wheeler-DeWitt equations (or quantum Hamiltonian constraint equations) can be interpreted to give a version of a many fingered time evolution of the quantum state
of geometry in the time $\tau$, in either the connection or the spin network representation.

In the following we restrict ourselves to the case where spacetime has a topology $\Sigma \times R$ where $\Sigma$ is a compact three manifold without boundary.  The case where there is a boundary or asymptotic conditions is interesting, but is reserved for a possible future paper.

\section{The Plebanski formulation of unimodular gravity}

Henneaux and Teitelboim\cite{HTuni} reformulated unimodular gravity so that the action depends on the full unconstrained 
metric and the gauge symmetry includes the full diffeomorphism group of the manifold. 
They do this as follows.  They introduce two auxiliary fields.  The first 
is a three form\footnote{We use greek indices for four dimensional manifold indices and latin indices for thee dimensional indices}  
$a_{\alpha \beta \gamma }$, whose field strength is $b_{\alpha \beta \gamma \delta}=da_{\alpha \beta \gamma \delta}$.  The dual is a density\footnote{We use the notation where tildes
refer to densities.}
\f
\tilde{b}= \frac{1}{4!} \epsilon^{\alpha \beta \gamma \delta} b_{\alpha \beta \gamma \delta}= \partial_\alpha \tilde{a}^\alpha
\ff
where $\tilde{a}^\alpha$ is the vector density field defined as 
$\tilde{a}^\alpha= \frac{1}{6} \epsilon^{\alpha \beta \gamma \delta}a_{ \beta \gamma \delta}$.
The second is scalar field $\phi$ which serves as a lagrange multiplier. 

Their action translates into the Plebanski formulation as 
\f
S^{HT} = \int_{\cal M} \left ( B^{i} \wedge F_{i} - \Phi_{ij} B^i \wedge B^j - \phi B_i \wedge B^i  + {\cal L}^{matter} \right )
+ \frac{1}{8\pi G}  \phi \tilde{b}
\label{HT}
\ff
where we impose the tracefree condition
\f
\Phi_{ii}=0
\ff
By varying $\phi$ we find the unimodular condition emerging as an equation of motion
\f
b= B_i \wedge B^i
\label{uni5}
\ff
Varying $\tilde{a}^a$ we find
\f
\partial_a \phi =0
\label{HT1}
\ff
so that the field $\phi$ becomes a spacetime constant on solutions, so we can write 
\f
\phi (x) = \Lambda
\ff
Varying $B^i $ we find the Einstein equations for any value of the constant $\Lambda$
\f
F_i = \Lambda B_i +\Phi_{ij} B^j 
\label{HT2}
\ff
plus matter terms.  The last equation of motion is
\f
B^i \wedge B^j - \frac{1}{3}\delta^{ij} B^i \wedge B_k=0
\ff

It makes sense to define the field
\f
\Phi^\prime_{ij}= \Phi_{ij} + \frac{1}{3} \delta_{ij} \phi
\ff
so that $\Phi^\prime_{ii}=\phi$.  If we use the equation of motion for $B^i$ we can rewrite the action as 
\f
S^{HT2}= \int_{\cal M} \left (   F^{i} \wedge F^{j} (\Phi^\prime)^{-1}_{ij} + {\cal L}^{matter} \right )
+ \frac{1}{8\pi G}  \phi \tilde{b}
\label{HT3}
\ff

Finally,  we can give an interpretation of the field $\tilde{a}$ as follows.  Let us integrate 
(\ref{uni5}) over a region of spacetime $\cal R$ bounded by two spacelike surfaces $\Sigma_1$ and $\Sigma_2$.  Then we have
\f
\int_{\Sigma_2} a - \int_{\Sigma_1} a = Vol = \int_{\cal R} \sqrt{-g}
\ff
That is $a$ pulled back into the surface is a time coordinate that measures the total four volume to the past of that surface.  We can consider that time coordinate associated to a surface $\Sigma$ to be 
\f
\tau= \int_{\Sigma} a  . 
\ff

\section{Constrained hamiltonian dynamics of Plebanski unimodular gravity}

It is easy to construct the constrained hamiltonian framework for the  Henneaux-Teiltelboim form of the theory (\ref{HT3}).  We do the usual $3+1$ decomposition and define momenta for all the fields.  For this and the next section we neglect matter, it is trivial to reinsert it.  We
find the canonical momenta for the gauge field
\f
\tilde{E}^a_i = \epsilon^{abc}F_{bc}  (\Phi^\prime)^{-1}_{ij} 
\label{tildeEdef}
\ff
and also the primary constraints 
\f
{\cal E}= \pi_0 - \phi =0
\label{Edef}
\ff
as well as 
\f
P_{\Phi}^{ij} =\Pi^0_i = \pi_\phi  =\pi_c=  0
\ff

Here $\pi_a, \pi_0$ are the momenta conjugate to the $\tilde{a}^a$ and 
$\tilde{a}^0$, respectively.    

The definition (\ref{Edef}) cannot be inverted to eliminate the velocities unless the following primary constraints are also imposed.
\f
{\cal D}_a = \tilde{E}^b_i F_{ab}^i =0 
\label{Ddef}
\ff
which generate diffeomorphisms and the modified Hamiltonian constraints
\f
{\cal H}= \epsilon^{ijk} \tilde{E}^a_i \tilde{E}^b_j F_{ab k} - \phi det ( \tilde{E}^a_i) =0  
\label{Hdef}
\ff

We also find from the preservation of the vanishing of $\Pi^0_i $ and $\pi_c $, respectively the two Gauss's law constraints
\f
{\cal G}^i = {\cal D}_a \tilde{E}^{ai} =0 
\label{Gdef}
\ff
\f
G_c= \partial_c \pi_0 =0
\label{Gcdef}
\ff 
where we have used (\ref{Edef}).

We can eliminate $\pi$ and $\pi_\phi$ by using (\ref{Edef}) with (\ref{Hdef}) to find
\f
\tilde{\cal W}= \epsilon^{ijk} \tilde{E}^a_i \tilde{E}^b_j F_{ab k} - \pi_0 det ( \tilde{E}^a_i) =0  
\label{Wdef}
\ff

The standard calculation shows that there is also a Hamiltonian, which is
\f
H= \int_\Sigma (\partial_a \tilde{a}^a ) \left ( \frac{\epsilon^{ijk} \tilde{E}^a_i \tilde{E}^b_j F_{ab k} }{det ( \tilde{E}^a_i) }   \right ) 
\label{hamdef}
\ff

If one wants one can alternatively eliminate (\ref{Gcdef}) and replace it by 
\f
S_c \equiv \partial_c \left    (   \frac{\epsilon^{ijk} \tilde{E}^a_i \tilde{E}^b_j F_{ab k} }{det ( \tilde{E}^a_i) }   \right ) =0
\label{Scdef}
\ff
Thus, the hamiltonian then vanishes on the constraint surface as (\ref{Scdef}) implies that
there is for every solution a constant $\Lambda$ for which  
\f
\epsilon^{ijk} \tilde{E}^a_i \tilde{E}^b_j F_{ab k} = \Lambda \cdot det( \tilde{E}^a_i). 
\ff
 The Hamiltonian is then on the constraint surface equal to
 \f
H= \Lambda \int_\Sigma (\partial_a \tilde{a}^a ) \approx 0
\label{ham2}
\ff

The constraints $\pi_a$ and ${\cal G_c}$ generate gauge transformations, respectfully 
\f
\tilde{a}^a \rightarrow \tilde{a}^a+ \tilde{r}^a , \ \ \ \ \ \ \tilde{a}^0 \rightarrow \tilde{a}^0+ \partial _c \tilde{s}^c 
\ff
where $\tilde{r}^a$ and $\tilde{s}^c$ are arbitrary vector densities. 

\section{Physical observables of the classical theory}

We first show that there is in this formalism a definite method for calculating physical observables which 
is no more complicated than solving evolution equations as in unconstrained dynamical theories.

Let us eliminate the pair  $(\tilde{a}^a , \pi_a)$ by solving the constraints $\pi_a=0$ at the same time we 
choose a gauge generated by that constraint in which $ \tilde{a}^a =0$.   This gives us 
a formulation of the theory on the phase space $\Gamma$ defined by canonical pairs  $(A_a^i, \tilde{E}^a_i )$,  and $(\tilde{a}^0 , \pi_0)$.
The constraints are $\cal W$, (\ref{Wdef}), ${\cal D}_a$ (\ref{Ddef}), ${\cal G}^i $(\ref{Gdef}), ${\cal G}_c $ (\ref{Gcdef}).  Let us note that
\f
\{ \tau , \pi_o (x) \} =1
\ff
and write $\cal W$ as the undensitized form
\f
{\cal W}= \pi_0 -h  =0  
\label{Wdef2}
\ff
where $h$ is 
\f
h=\frac{\epsilon^{ijk} \tilde{E}^a_i \tilde{E}^b_j F_{ab k} }{det ( \tilde{E}^a_i)}
\ff
Let us now consider a physical observable on $\Gamma$
\f
{\cal O} = {\cal O}[ A, E, \tilde{a}^0, \pi_0] 
\ff
Let us assume that it is locally gauge invariant and spatially diffeomorphism invariant so it satisfies
\f
\{ {\cal G}^i , {\cal O} \}=0
\ff
\f
\{ {\cal D}(v) , {\cal O} \}=0
\label{Deq}
\ff
for any vector field $v^a$ on $\Sigma$.   We want to solve 
\f
\{ {\cal W} , {\cal O} \}=0
\label{Weq}
\ff
\f
\{ {\cal G}_a , {\cal O} \}=0
\label{Geq}
\ff
We solve the latter, (\ref{Geq}) first.  This gives us
\f
\frac{\partial}{\partial x^a}\left (  \frac{\partial {\cal O}}{\partial \tilde{a}^0(x)} \right ) =0
\ff
This is solved by making $\cal O$ a function only of the integral $\tau =\int_\Sigma \tilde{a}^0$.  Now
we can consider (\ref{Weq}).  This is one equation for every point $x \in \Sigma$, 
\f
0= \{ {\cal W}(x)  , {\cal O}[\tau ] \}= \frac{\partial {\cal O}[\tau ]}{\partial \tau } - \{{ \cal O}[\tau ], h (x) \}
\label{Weq2}
\ff
However it is straightforward to show that if this is satisfied at any one point $x\in \Sigma$ it is satisfied
for all points $y \in \Sigma$.  Consider that $x$ and $y$ are nearby  so that there is a 
vector field $v^a$ such that $h(y) = h(x) + \{ {\cal D}(v) , h(x) \}+...$, where the $...$ indicate
higher order terms in brackets with ${\cal D}(v)$ that come from exponentiation.   Then we can write
\begin{eqnarray}
 \{{ \cal O}[\tau ], h (y) \} &=&  \{{ \cal O}[\tau ], h (x) \} +  \{{ \cal O}[\tau ], \{ {\cal D}(v) , h(x) \} \} +... \nonumber \\
 &=&  \{{ \cal O}[\tau ], h (x) \} -  \{ {\cal D}(v), \{   h(x), { \cal O}[\tau ]\} \} - 
   \{h(x) , \{    { \cal O}[\tau ], {\cal D}(v)\} \}  \nonumber \\
   &=&  \{{ \cal O}[\tau ], h (x) \} 
\end{eqnarray}
where $ \{ {\cal D}(v), \{   h(x), { \cal O}[\tau ]\} \} $ vanishes by (\ref{Weq2}) and 
$\{    { \cal O}[\tau ], {\cal D}(v)\}$ vanishes by (\ref{Deq}). 

Thus we conclude that a complete set of conditions on ${\cal O}[\tau ]$ are the constraint equations
(\ref{Gdef}), (\ref{Ddef}) together with the condition that it is a function of $\tilde{a}^0$ only through
a dependence on the physical time, $\tau$ together with the evolution equation
\f
\frac{\partial {\cal O}[\tau ]}{\partial \tau } = \{{ \cal O}[\tau ], h (x) \}
\ff
for any $x$.

\section{Construction of the path integral quantization}

The canonical theory we have arrived at has canonical pairs,
 $(A_a^i, \tilde{E}^a_i )$, $(\tilde{a}^a , \pi_a)$  and $(\tilde{a}^0 , \pi_0)$.   The constraints
are $\cal W$, (\ref{Wdef}), ${\cal D}_a$ (\ref{Ddef}), ${\cal G}^i $(\ref{Gdef}), ${\cal G}_c $ (\ref{Gcdef})
and $\pi_c$.  The ${\cal G}_c $ count as 
one constraint because 
\f
\partial_{[a} {\cal G}_{c]} = 0
\ff

Associated to these $11$ constraints are $11$ gauge fixing conditions.  The first seven are standard from the Ashtekar formulation
of general relativity, the four new ones gauge fix the $\tilde{a}^a$ and  $\tilde{a}^0$ and will be discussed below.

There is also a Hamiltonian which is (\ref{hamdef}) which is non-vanishing on the constraint surface.

We follow the standard construct the path integral from the the gauge fixed Hamiltonian quantum dynamics.  The partition 
function is
\begin{eqnarray}
Z &=&  \int  dA_a^i d\tilde{E}^a_i d\tilde{a}^0 d\tilde{a}^a d\pi_a d\pi_o \delta ({\cal W})   \delta ({\cal D}_a ) \delta ({\cal G}^i ) \delta ({\cal G}_c) \delta (\pi_c ) 
\delta (\mbox{ gauge fixing})  Det_{FP}
\nonumber \\
 & &\times  exp \  \imath \int dt \int_\Sigma \left (  \tilde{E}^a_i \dot{A}_a^i + \pi_0 \dot{\tilde{a}}^0  + \pi_c \dot{\tilde{a}}^c  -  (\partial_a \tilde{a}^a ) \left ( \frac{\epsilon^{ijk} \tilde{E}^a_i \tilde{E}^b_j F_{ab k} }{det ( \tilde{E}^a_i) }   \right ) 
  \right ) 
 \label{partition1}
\end{eqnarray}

The first thing we do is to integrate over $\pi_0$.  This uses up the delta function in ${\cal W}$, implying the substition
\f
\pi_0 \rightarrow \left    (   \frac{\epsilon^{ijk} \tilde{E}^a_i \tilde{E}^b_j F_{ab k} }{det ( \tilde{E}^a_i) }   \right ) 
\ff
in the rest of the integrand.  This turns the delta functional in ${\cal G}_c$ into one in ${\cal S}_c$ so the partition function is now
\begin{eqnarray}
Z &=&  \int  dA_a^i d\tilde{E}^a_i d\tilde{a}^0  d\tilde{a}^a d\pi_c    \delta ({\cal D}_a ) \delta ({\cal G}^i ) \delta ({\cal S}_c)  \delta (\pi_c ) 
\delta (\mbox{ gauge fixing})  Det_{FP}
\nonumber \\
 & &\times  exp \  \imath \int dt \int_\Sigma \left (  \tilde{E}^a_i \dot{A}_a^i +  \left    (   \frac{\epsilon^{ijk} \tilde{E}^a_i \tilde{E}^b_j F_{ab k} }{det ( \tilde{E}^a_i) }   \right ) ( \dot{\tilde{a}}^0 +\partial_c \tilde{a}^c )    + \pi_c \dot{\tilde{a}}^c   \right ) 
 \label{partition2}
\end{eqnarray}

We now introduce a vector density $\tilde{w}^a$ to exponentiate the constraint ${\cal S}_a$, and 
integrate over the $\pi_c$. yielding 
\begin{eqnarray}
Z &=&  \int  dA_a^i d\tilde{E}^a_i d\tilde{a}^0   d\tilde{a}^a d\tilde{w}^c  \delta ({\cal D}_a ) \delta ({\cal G}^i ) 
\delta (\mbox{ gauge fixing})  Det_{FP}
\nonumber \\
 & &\times  exp \  \imath \int dt \int_\Sigma \left (  \tilde{E}^a_i \dot{A}_a^i   +  \left    (   \frac{\epsilon^{ijk} \tilde{E}^a_i \tilde{E}^b_j F_{ab k} }{det ( \tilde{E}^a_i) }   \right ) ( \dot{\tilde{a}}^0  +\partial_c \tilde{a}^c  + \partial_c \tilde{w}^c )  \right ) 
 \label{partition3}
\end{eqnarray}

We now may shift the definition of $\tilde{a}^a$ by
\f
\tilde{a}^a \rightarrow \tilde{a}^a - \tilde{w}^a
\ff
This eliminates any dependence on $\tilde{w}^a$ in the integrand so we trivially do the integral over
$d\tilde{w}^a$.   
At the same time we exponentiate the two remaining constraints, ${\cal D}_a$ and ${\cal G}^i$ with lagrange multipliers,
respectively, $N^a$, the lapse and $A_0^i$.  We then have 
\begin{eqnarray}
Z &=&  \int  dA_a^i  dA_0 ^i d\tilde{E}^a_i  dN^a d\tilde{a}^0  d\tilde{a}^c  
\delta (\mbox{ gauge fixing})  Det_{FP}
\nonumber \\
 & &\times  exp \  \imath \int dt \int_\Sigma \left (  \tilde{E}^a_i \dot{A}_a^i  +  \left    (   \frac{\epsilon^{ijk} \tilde{E}^a_i \tilde{E}^b_j F_{ab k} }{det ( \tilde{E}^a_i) }   \right ) ( \dot{\tilde{a}}^0   + \partial_c \tilde{a}^c )  +N^a {\cal D}_a + A_0^i {\cal G}^i \right ) 
 \label{partition4}
\end{eqnarray}

We now can introduce the lapse $N$, which is an density of weight minus unity, by means of an insertion of unity, in the form
\f
1= \int dN \delta \left ( N-\frac{\dot{\tilde{a}}^0 +\partial_c \tilde{a}^c }{Det(\tilde{E}^{ai})} \right )
\ff
This gives us 
\begin{eqnarray}
Z &=&  \int  dA_a^i  dA_0 ^i d\tilde{E}^a_i  dN^a dN d\tilde{a}^0  d\tilde{a}^c  \delta \left (N-\frac{\dot{\tilde{a}}^0 +\partial_c \tilde{a}^c }{Det(\tilde{E}^{ai})} \right )
\delta (\mbox{ gauge fixing})  Det_{FP}
\nonumber \\
 & &\times  exp \  \imath \int dt \int_\Sigma \left  (  \tilde{E}^a_i \dot{A}_a^i  +   N \epsilon^{ijk} \tilde{E}^a_i \tilde{E}^b_j F_{ab k}    +N^a {\cal D}_a + A_0^i {\cal G}^i   \right  ) 
 \label{partition5}
\end{eqnarray}

We can change variables from $\tilde{E}^{ai}, N $ and $N^a$ back to the frame fields $e^\mu$ so we have 
\begin{eqnarray}
Z &=&  \int  dA_a^i  dA_0 ^i de^\mu d\tilde{a}^0 d\tilde{a}^c  \delta \left (det(e) -\dot{\tilde{a}}^0 +\partial_c \tilde{a}^c \right )
\delta (\mbox{ gauge fixing})  Det_{FP}
\nonumber \\
 & &\times  exp \  \imath \int dt \int_\Sigma \left  (  e^\mu \wedge e^\nu \wedge F_{\mu \nu}^+    \right  ) 
 \label{partition6}
\end{eqnarray}

\subsection{Gauge fixing}

We now can introduce the four gauge fixing conditions  
\f
\tilde{f}_0 = \tilde{a}^0 - t (\epsilon_0 - \partial_c \tilde{a}^c ) =0, \ \ \ \  \tilde{f}^c= \tilde{a}^c =0
\ff
where  $t$ is the time coordinate.  $\epsilon_0$ is a fixed density needed to make the 
density weights balence out. Then we have  
\f
\dot{\tilde{a}}^0 +\partial_c \tilde{a}^c=\epsilon_0.  
\ff
The path integral now becomes
\begin{eqnarray}
Z &=&  \int  dA_\mu^i  de^\mu  \delta \left (det(e) -\epsilon_0  \right )
\delta (\mbox{ gauge fixing})  Det_{FP}^\prime
\nonumber \\
 & &\times  exp \  \imath \int dt \int_\Sigma \left  (  e^\mu \wedge e^\nu \wedge F_{\mu \nu}^+    \right  ) 
 \label{partition7}
\end{eqnarray}
so we return to an action of the form of unimodular gravity.

From here we can work back to the Plebanski formalism following the usual steps, giving us
\begin{eqnarray}
Z &=&  \int  dA^i  dB^i d\Phi_{ij}   \delta \left ( B^i \wedge B_i -\epsilon_0  \right )\delta (\Phi_{ii} )
\delta (\mbox{ gauge fixing})  Det_{FP}^\prime
\nonumber \\
 & &\times  exp \  \imath \int dt \int_\Sigma \left  ( B^i \wedge F_i - \Phi_{ij} B^i \wedge B^j     \right  ) 
 \label{partition8}
\end{eqnarray}
Thus, the path integral is the same as is usually taken to generate spin foam models, with the
additional constraint that $B^i \wedge B_i$ is everywhere fixed to a background density $\epsilon_0$.

\section{The unimodular loop representation}

Unimodular gravity in the Ashtekar variables was studied by \cite{bombelliuni}, which we take as a starting point.  Their work
was done at a time when the roles of the volume operator and spin network basis were not fully appreciated, 
so it should be now possible to take their results some steps further.  Here we just make a few remarks to motivate further work.

\subsection{The connection representation}

To quantize in the loop representation we may begin with an extended connection representation
where the configuration space is a functional of connections $A^i$ on a three manifold $\Sigma$ plus
a scalar density field $\tilde{a}^0$.  (As $\tilde{a}^a$ plays no role we will eliminate it and its conjugate
momenta, which is a constraint.) For convenience we will rename $\tilde{a}^0$, $\tilde{T}$ and recall
that $\tau = \int_\sigma \tilde{T}$ measures total four volume since the beginning of the universe.  

Wavefunctionals are then functionals $\Psi (A, \tilde{T})$.   The diffeomorphism and gauge constraints
affect only the $A$ dependence in the usual way.  in addition,  there is the hamiltonian constraint (\ref{Hdef})
which may be written, in singly densitized form
\f
\imath \hbar \frac{\partial}{\partial \tilde{T}} \hat{det(e)} \Psi (A, \tilde{T})= \hat{\tilde{h}} \Psi (A, \tilde{T})
\ff
where $\tilde{h}$ is the singly densitized form of the hamiltonian constraint
\f
\tilde{h} = \frac{1}{\sqrt{q}}\epsilon^{ijk} \tilde{E}^a_i \tilde{E}^b_j F_{ab k} 
\ff

There is the additional constraint ${\cal G}_c$ (\ref{Gcdef}), which may be written.
\f
\imath \hbar \partial_c \frac{\partial}{\partial \tilde{T}}  \Psi (A, \tilde{T})=0 
\ff
This is solved by writing,
\f
 \Psi (A, \tilde{T})=  \Psi (A, \tau (\tilde{T}) )
\ff

Alternatively we can separate variables, by writing 
\f
 \Psi (A, \tilde{T})= \int d\lambda (x)  \Psi (A, \lambda) e^{\imath \int_\Sigma \tilde{T} \lambda (x) }
\ff
which is a functional integral over values of a  function $\lambda (x) $ on $\Sigma$.  We then find that locally
\f
 \hat{\tilde{h}} (x) \Psi (A, \lambda ) = \lambda (x)  \hat{det(e)} \Psi (A,\lambda )
 \label{localeq}
\ff
together with the condition that
\f
(\partial_c \lambda) \Psi (A, \lambda ) =0
\label{Gconstraint1}
\ff
We see that $\Psi (A,\lambda ) $ has support only on configurations where $\lambda $ is constant.  

We can integrate (\ref{localeq}) over any region ${\cal R}$ of the spatial manifold to find 
\f
 \hat{\tilde{h}}_{\cal R} \Psi (A, \lambda ) = \lambda   \hat{V}_{\cal R} \Psi (A,\lambda )
 \label{localeq2}
\ff
where 
\f
h_ {{\cal R}} = \int_{{\cal R}} \tilde{h} 
\ff
and 
\f
\hat{V}_ {{\cal R}} =  \int_{{\cal R}} \hat{\sqrt{q}} 
\ff

If we define
that constant value $\lambda (x) =\Lambda$ we find that the wavefunction evolves as
\f
 \Psi (A, \tau)= \int d\Lambda  \Psi (A, \Lambda) e^{\imath \tau \Lambda} 
\ff
This is now an ordinary integral over $\Lambda$.  

It is amusing to note that the Kodama state\cite{kodama} $\Psi_k (A, \lambda ) = e^{\frac{3}{\lambda}\int Y_{CS} (A) }$ is still a solution to (\ref{localeq}), with the state considered a function of variables
$A$ and $\lambda$.  With the standard point split regularizations, it solves the ordering
\f
\epsilon^{ijk} \hat{\tilde{E}}^a_i \hat{\tilde{E}}^b_j \left ( \hat{F}_{ab k} -\frac{\lambda}{3} \epsilon_{abc} \hat{\tilde{E}}^c_k    \right )  \Psi (A,\lambda )=0
\ff
Whether this offers any improvement of the interpretational issues facing the Kodama state is unclear.  

\subsection{The spin network representation}

We may now transform to the spin network representation, it is convenient to use the fact that 
in (\ref{Hdef}), $\pi_0$ multiplies $det{E}$ so we only need to have values of $\tilde{T}$ on 
vertices of valence four or higher, on which the volume operator doesn't vanish.  
Let ${\cal R}_i$ be a decomposition of $\Sigma$ into regions, each enclosing a single vertex $v_i$ of a spin network state of
valence four or higher.  Then we can associate to each vertex $v_i$, the integral
\f
\int_{{\cal R}_i} \tilde{T} = \tau_i
\ff
so that $\sum_i \tau_i =\tau$ which measures the past four-volume of the slice.  Note that the partition 
of $\tau$ so defined on vertex labels is arbitrary, as the regions are arbitrary parts of a regularization 
procedure.  

We then augment the usual definition of spin network states with these labels $\tau_i$ on each 
vertex of non-vanishing volume.  The quantum states are then functionals
\f
\Psi (\Gamma , \tau_i )
\ff
But the solutions to (\ref{Gcdef}) impose that all the $\tau_i =\tau$ so we have quantum states of the form
\f
\Psi (\Gamma , \tau )
\ff
that satisfy for every region $\cal R$,
\f
\imath \hbar \frac{\partial}{\partial \tau} \hat{V}_{\cal R}  \Psi (\Gamma , \tau ) =  \hat{h}_{\cal R}    \Psi (\Gamma , \tau )
\label{Hdef3}
\ff

\section{Conclusions}

In this paper we have extended to the Ashtekar and Plebanski formulations of gravity the results of \cite{me-uni},  in which it is shown that the quantum
effective action of unimodular gravity retains the unimodular property, (\ref{modify}).  The main results of this paper are the formal expression for the 
path integral in unimodular form of the Plebanski action, eq, (\ref{partition8}) and the unimodular form 
of the Wheeler-deWitt equation in spin network representation, eq. (\ref{Hdef3}).  We note that the when the region $\cal R$ in the latter is chosen to be the whole spatial manifold, we have a kind of Schroedinger equation.  But we should caution that this equation has to be satisfied for every region,  $\cal R$, of the manifold.   In particular, note that there are generically regions in spin network states in the kernel of the volume operator where the regularized forms of $\hat{h}_{\cal R}$ so far
studied act non-trivially.  These include nodes of valence three; on these the hamiltonian constraint will take its usual form.  What is new is the action on regions, such as those containing generic nodes of valence four or higher,  on which the action of both $\hat{V}_{\cal R}$ and $\hat{h}_{\cal R}$ is non-trivial.
The development of solutions in these regions will give something like a many fingered time, where the ``extent"  of each finger is determined by the action of the volume operator. 

Hence,  while unimodular gravity does offer a physical time coordinate, evolution in terms of which is defined by a Schroedinger equation, this does not get us out of the challenge of solving an infinite number of dynamical equations, one for each point or region of the spatial manifold. 

Given that there has been recently a great deal of progress in the construction and analysis of spin foam models, it would be very interesting to see if
a concrete definition can be given in spin foam terms of the path integral (\ref{partition8}).  We note that some progress in this direction has been made in 
\cite{marc} which also studies the unimodular form of loop quantum cosmology.  

\section*{ACKNOWLEDGEMENTS}

I am grateful to Niayesh Afshordi, Giovanni Amelino-Camelia, Laurent Freidel, Sabine Hossenfelder and Carlo Rovelli for crucial comments, as well as to Rafael Sorkin for conversations and  to Marc Geiller and Marc Henneaux for correspondence.  Research at Perimeter Institute for Theoretical Physics is supported in part by the Government of Canada through NSERC and by the Province of
Ontario through MRI.

\end{document}